\documentclass[12pt]{article}
\usepackage[top=1.5cm, bottom=1.5cm, left=1.5cm, right=1.5cm]{geometry}
\usepackage{longtable}
\usepackage{amsmath}
\usepackage{txfonts}
\usepackage{graphicx}
\usepackage{amssymb}
\usepackage[labelfont=bf]{caption}
\begin{document}
\title{Orbital Period Changes and Their Explanations in Two Algol Binaries: WY Per and RW Leo  }
\author{L\textsc{iao} Wen-Ping $^{1,2,3,4}$, Q\textsc{ian} Sheng-Bang $^{1,2,3}$}

\footnotetext[1]{\scriptsize{National Astronomical
Observatories/Yunnan Observatory, Chinese Academy of Sciences,\\
P.O. Box 110, 650011 Kunming, P. R. China}}

\footnotetext[2]{\scriptsize{Key Laboratory for the Structure and
Evolution of Celestial Objects, Chinese Academy of Sciences.}}

\footnotetext[3]{\scriptsize{Graduate School of the CAS, 100049,
Beijing, P.R. China/Yuquan Road 19\# , \\SijingShan Block, 100049,
Beijing City, P. R. China.}}

\footnotetext[4]{\scriptsize{E-mail: liaowp@ynao.ac.cn}}


\maketitle


\begin{abstract}
\small

\noindent{WY Per and RW Leo are two Algol-type binaries. Based on
our new CCD observations and the almost century-long historical
record of the times of primary eclipse for WY Per and RW Leo, the
orbital period changes and their explanations were reanalyzed and
rediscussed in detail. It is found that the orbital period of WY Per
shows a cyclic oscillation with a period of $P_{3} = 71.5$ yr and a
semiamplitude of $A_{3} = 0^{\textrm{d}}.0739$. The period variation
can be interpreted by the light-travel time effect (LTTE) via the
presence of a third body in an eccentric orbit with an eccentricity
of $e_{3} \simeq 0.602$ in the system. For RW Leo, its orbital
period shows complex variations. Two cyclic variations (i.e., $P_{3}
= 77.8$ yr and $A_{3} = 0^{\textrm{d}}.033$, and $P_{4} = 39.1$ yr
and $A_{4} = 0^{\textrm{d}}.022$) are discovered. The cyclic
variation of $P_{3} = 77.8$ yr and $A_{3} = 0^{\textrm{d}}.033$ can
be attributed to the LTTE via the presence of a third body in an
eccentric orbit with an eccentricity of $e_{3} \simeq 0.732$.}

\end{abstract}


\begin{bfseries}
\noindent{stars: binaries: close --- stars: binaries: eclipsing
--- stars: individuals (WY Persei, RW Leonis) --- stars: multiple}
\end{bfseries}




\section{Introduction}

Cyclical period changes of stellar eclipsing binary systems can be
investigated by analyzing the $(O-C)$ diagram showing the difference
between the observed times of light minimum and those computed with
a given ephemeris. Such cyclical period changes are a fairly common
phenomenon in close binary systems and are usually explained as due
to either the magnetic activity of one or both components (e.g.,
Applegate 1992) or to the light-travel time effect(LTTE) of a third
body. The Applegate mechanism was supported by the observational
fact deduced from the period changes of 101 Algols that all cases of
cyclical period changes are restricted to binaries with spectral
types of the secondary later than F5, and no cyclic variations were
detected for systems with spectral types earlier than F5 (Hall
1989). However, in our recent compilation of the available
information on orbital period variations for different classes of
close binaries (Liao \& Qian 2010), the cyclic period changes are
also discovered in early-type binaries, which can be plausibly
interpreted by the presence of an additional body. Additionally, as
discussed by Qian et al. (2008a) and Lanza (2006), the Applegate
mechanism is not adequate to explain the orbital period modulation.
Therefore, the most plausible explanation of the cyclical period
changes is the LTTE via the presence of a third body (Liao \& Qian
2010). In the present paper, we will analyze the cyclical period
changes of two Algol-type binaries WY Per and RW Leo derived from
the long historical record of the times of light minimum and explore
the causes of these period changes.

WY Per (=AN 14.1917, $V_{max}$ = 11.5 mag) is an Algol-type binary.
It was discovered to be a variable by Wolf (1917). The first several
times of light minimum come from 1921 derived by Hoffmeister (1921)
by using visual observations, and he also gave the first linear
elements. From then on, there were dozens of visual and several
photographic and CCD observations obtained (see table 1). Two new
CCD observations were obtained by using the two telescopes at the
Yunnan Observatory in the present paper. Afterwards, the variation
of period for WY Per was pointed out by Whitney (1959), but at that
time the LTTE can not identified conclusively, Wolf et al. (2004)
discussed LTTE of the system and gave the parameters of third body.
Brancewicz \& Dworak (1980) computed the geometric and physical
parameters for components of this eclipsing binary stars. The
spectral types were classified as A0+K0IV in Budding (1984),
afterwards, Svechnikov and Kuznetsova (1990) gave A0+K2.5IV. With
more times of light minimum, we reanalyze the orbital period of WY
Per that reveals the appearance of a massive third body in the
system.

RW Leo(=AN 24.1914, $V_{max}$ = 11.9 mag) is also an Algol-type
binary. The first several times of light minimum also come from 1921
derived by Hoffmeister (1921), and he gave the first linear
elements. Afterwards, four photographic times of light minimum, for
RW Leo, were published by Whitney (1959), and he stated that the
period is variable. RW Leo was referred to as a photoelectrically
neglected eclipsing binary by Koch et al. (1979). Spectroscopic
observations of the system have been published by Halbedel (1984)
and by Kaitchuck (1985). The first photoelectric UBV light curves
were reported by Walker (1992), who pointed out that this binary is
a typical semi-detached system consisting of an A3V primary and an
F4 subgiant secondary. By using 29 visual and photographic times of
light minimum (only one CCD observation), Qian (2003) first gave
orbital period study of RW Leo, a cyclic oscillation of 38.4 yr was
discovered to superimpose on secular period increase with $dP/dt =
+1.78 \times 10^{-7} \textrm{d yr}^{-1}$. As new and more accurate
observational material has accumulated since then, in the present
work, with 70 times of light minimum, the detailed orbital period
changes of RW Leo are reinvestigated by us.
\section{ CCD Photometry of WY Per and RW Leo}
CCD photometry for WY Per was carried out on Feb 15, 2007 and Dec
29, 2009 with two DW 436 CCD cameras attached to the 1.0- and 0.6-m
telescopes at the Yunnan Observatory, respectively. The effective
field of view of the photometric system is about $6^{\prime}.5\times
6^{\prime}.5$ for 1.0-m and $12^{\prime}.5\times 12^{\prime}.5$ for
0.6-m at the Cassegrain focus. The R filter, close to the standard
Johnson UBVRI system, was used. The integration time is 30 s (1.0-m)
and 100 s (0.6-m) for each image, respectively. The comparison and
check stars are GSC 02870-01440 ($03^{h}38^{m}28^{s}.93$,
$+42^{\circ}40^{\prime}24.7^{\prime\prime}$) and
($03^{h}38^{m}29^{s}.90$ ,
$+42^{\circ}41^{\prime}04.5^{\prime\prime}$). The PHOT task of IRAF,
which measures the aperture magnitude for a list of stars, was used
to reduce the observed images. By using our photometric data, two
times of light minimum, HJD 2454147.1184($\pm0.0002$) and HJD
2455195.13297($\pm0.00030$), were determined.

RW Leo was monitored on Jan 22, 2008 and Jan 18, 2010 with the 1.0-m
telescope at the Yunnan Observatory. The R filter, close to the
standard Johnson UBVRI system, was used. The integration time is 90
s for the first night and 60 s for the second night. The comparison
and check stars are GSC 00839-00530 ($10^{h}39^{m}18^{s}.43$,
$+09^{\circ}00^{\prime}56.7^{\prime\prime}$) and
($10^{h}39^{m}22^{s}.15$,
$+09^{\circ}02^{\prime}22.9^{\prime\prime}$). By using our
photometric data, two times of light minimum, HJD
2454488.38902($\pm0.00027$) and HJD 2455215.24169($\pm0.00008$),
were determined.

\section[]{Orbital Period Changes for Two Algol-type Binaries WY Per and RW Leo}
\subsection[]{Period Analysis for WY Per}

Whitney (1959) pointed out that the period of WY Per is variable,
Wolf et al. (2004) discussed LTTE of the system and gave the
parameters of third body. In order to determine more accurate
parameters of LTTE with more times of light minimum, we collected
all times of light minimum which cover more than 90 years(i.e., 59
visual, 4 photographic, and 10 CCD observations). They are listed in
the first column of table 1. Using the linear ephemeris formula
given by Kreiner et al. (2001),
\begin{eqnarray}
MinI=2446002.3497+3^{d}.327123\times{E},
\end{eqnarray}
the epoch number ($E$) and $(O-C)_1$ values of WY Per were
calculated, which are listed in the fourth and fifth column of table
1, and $(O-C)_1$ values vs. the epoch number is plotted in the upper
panel of figure 1, where open circles refer to photographic or
visual observations, filled circles to CCD ones. As displayed in the
upper panel of figure 1, the period change of the binary is complex.
A simple sinusoidal variation cannot fit all of the data
satisfactorily. To fit the general $(O-C)_1$ trend satisfactorily,
we consider a more general case of combining a cyclic variation with
an eccentricity and a long-term change of period. Weights of 1 and 8
were assigned to lower-precision observations (photographic or
visual ones) and high-precision observations (CCD or photoelectric
ones), respectively. With weights of a nonlinear fitting, the
following equation is formed to fit $(O-C)_1$ curve,
\begin{eqnarray}
{\rm O-C}&=&\Delta JD_{0}+ \Delta P_{0}E +\frac{\beta}{2}E^{2} + A[(1-e_{3}^{2})\frac{\sin(\nu_{3}+\omega_{3})}{1+e_{3}\cos\nu_{3}}+e_{3}\sin\omega_{3}]\nonumber\\
   &=&\Delta JD_{0}+\Delta P_{0}E +\frac{\beta}{2}E^{2}+A[\sqrt{1-e_{3}^{2}}\sin E_{3}\cos \omega_{3}+\cos E_{3}\sin
\omega_{3}],
\end{eqnarray}
\noindent where $E$ is the epoch number, $\Delta JD_{0}$ and $\Delta
P_{0}$ are the correction values for the initial epoch and orbital
period of the binary, $\beta$ is the long-term change of the orbital
period(day $\textrm{cycle}^{-1}$); $A=a_{12}\sin i_{3}/c$,
$a_{12}\sin i_{3}$ is the projected semimajor axis, $c$ the speed of
light; $e_{3}$ the eccentricity of a supposed third body, $\nu_{3}$
the true anomaly of the position of the eclipsing pair's mass centre
on the orbit, $\omega_{3}$ the longitude of the periastron of the
eclipsing pair's orbit around the third body, and $E_{3}$ is the
eccentric anomaly.

The Kepler equation provides the connection between the eccentric
anomaly ($E_{3}$) and the observed times of light minimum:
\begin{eqnarray}
M_{3}&=&E_{3} - e_{3} \sin E_{3}=\frac{2\pi}{P_{3}} (t - T_{3}),
\end{eqnarray}
\noindent where $M_{3}$ is the mean anomaly, $T_{3}$ the time of
periastron passage, $P_{3}$ the period of a supposed third body, and
$t$ is the observed times of light minimum. It is clear from
equations (2) and (3) that we could determine five parameters
($P_{3}$, $T_{3}$, $A$, $\omega_{3}$, $e_{3}$) so that fit the
$(O-C)_1$ trend. For the eccentric anomaly ($E_{3}$) and the true
anomaly($\nu_{3}$) are not mutually absolute parameters, here we
expand $E_{3}$ approximatively with Bessel's series. Then
Levenberg-Marquardt method is adopted by using equations (2) and (3)
to a nonlinear fit of $(O-C)_1$. During the fit of $(O-C)_1$ curve
for WY Per, $\beta$ was approximately assumed to be zero due to the
very large error of it. The fitting parameters of the third body
from our analysis are listed in the upper part of table 3. It is
clear from the table 3 that our parameters of the orbit of the third
body for WY Per differ from those determined by Wolf et al. (2004).
It is found that the orbital period of WY Per shows a cyclic
oscillation with a period of 71.5 yr and a semiamplitude of 0.0739
days, which is more easily seen from the middle panel of figure 1,
where the linear correction part of equation (2) was subtracted to
the $(O-C)_1$ values. The residuals from the whole effect are
displayed in the lower panel of figure 1 and listed in the seventh
column of table 1, where no regularity can be found indicating that
our calculation gives a good fit to the $(O-C)_{1}$ diagram.

\subsection[]{Period Analysis for RW Leo}
For the Algol-type binary RW Leo, Whitney (1959) stated that the
period is variable. Qian (2003) first studied the period changes of
this binary with only 29 times of light minimum, but since then
detailed period analysis of RW Leo has not been formed. In order to
understand the character of the period variation of this binary, a
total of 70 times light minimum cover more than 90 years were
collected, which are listed in the first column of table 2. With the
linear ephemeris formula of Kreiner et al. (2001),
\begin{eqnarray}
MinI=2443324.7374+1^{d}.68254017\times{E},
\end{eqnarray}
the $(O-C)_1$ values, listed in the fifth column of table 2, were
calculated. The corresponding $(O-C)_1$ curve is displayed in the
upper panel of figure 2, where open circles refer to photographic or
visual observations, filled circles refer to CCD ones. As shown in
the upper panel of figure 2, one may think the general trend of
$(O-C)_1$ changes in a simple way, i.e., single, two, or even three
periodic terms are almost sufficient to fit the $(O-C)_1$ curve.
However, after trying several fitting patterns, we found a simple
single, two, or even three periodic terms cannot fit all of the data
satisfactorily. Finally, we used a linear-plus-two cyclic
variations(one has eccentric orbit) ephemeris to fit the trend of
$(O-C)_1$ curve. Therefore, we add a sinusoidal term
$A_{4}\sin(\omega_{4} E-\phi)$ to the equation (2) to fit the
$(O-C)_1$ curve, and the same weighting scheme with WY Per is
applied. $\beta$ was also approximately assumed to be zero due to
the very large error of it,  More times of light minimum are
required in the future for determining the accurate value of
long-term period change. The corresponding fitting parameters of the
orbital solutions are listed in the upper part of table 4. It is
found that the orbital period of RW Leo shows two cyclic
oscillations (i.e., $P_{3} = 77.8$ yr and $A_{3} =
0^{\textrm{d}}.033$, and $P_{4} = 39.1$ yr and $A_{4} =
0^{\textrm{d}}.022$). The solid line in figure 2 represents the
combination of a linear ephemeris and two cyclic oscillations. After
the linear ephemeris and simple sinusoidal variation were removed,
the $(O-C)_2$ values are displayed in the middle panel of figure 2,
where the cyclic oscillation with an eccentricity of $e_{3} \simeq
0.732$ can be seen more clearly. The $(O-C)_3$ values, which are
plotted in the lower panel of figure 2, are the ones after
subtracting the linear ephemeris and the eccentric orbit
oscillation. The residuals from fit are displayed in the figure 3
and listed in the eighth column of table 2. To detect possible
regular trends in the residuals plotted in figure 3, more
high-precision times of light minimum are needed from future
observations.
\section[]{DISCUSSIONS AND CONCLUSIONS}
\subsection[]{Explanations of the Cyclic Period Changes in WY Per}
For WY Per, secondary component's spectral type is later than F5
(K2.5IV), this cyclic variation can be explained as either the
magnetic activity cycles in one or both components (Applegate 1992),
or the light-travel time effect (LTTE) via the presence of a third
body (Frieboes-Conde \& Herczeg 1973; Chambliss 1992; Borkovits \&
Heged\"{u}s 1996). Using the following equation (Rovithis-Livaniou
et al. 2000),
\begin{eqnarray}
 \Delta P=\sqrt{2[1-\cos(2\pi P/P_{3})]} \times A ,
\end{eqnarray}
\noindent the amplitude of the period oscillation can be computed.
The rate of the period variation $\Delta P$/ $P$ can be used to
calculate the variation of the quadruple moment $\Delta Q$ required
to reproduce this cyclic change of WY Per, with the following
equation (Lanza \& Rodon\`{o} 2002)
\begin{eqnarray}
 \frac{{\Delta P}}{P}=-9\frac{\Delta Q}{Ma^{2}},
\end{eqnarray}
\noindent $\Delta Q_{1}$ = 1.44 $\times 10^{50}$ g $\textrm{cm}^{2}$
and $\Delta Q_{2}$ = 6.41 $\times 10^{49}$ g $\textrm{cm}^{2}$ for
both components in the binary are estimated, respectively, which are
not well within the limits for the close binary ($10^{51}-10^{52}$ g
$\textrm{cm}^{2}$, Lanza \& Rodon\`{o} (1999)) and therefore the
cyclical period variation in WY Per could not interpreted by the
mechanism of Applegate.

Therefore, the cyclical period change in WY Per is explained as LTTE
arises from the gravitational influence of a third body. With the
same masses as Wolf et al. (2004) used ($m_{1}=2.25M_{\odot}$,
$m_{2}=1.0M_{\odot}$) and the known equation,
\begin{eqnarray}
f(m) =\frac{(m_{3}\sin{i_{3}})^{3}}
{(m_{1}+m_{2}+m_{3})^{2}}=\frac{4\pi^{2}}{G{P_{3}}^{2}}\times(a_{12}\sin{i_{3}})^{3},
\end{eqnarray}
\noindent where $m_{1}$, $m_{2}$ and $m_{3}$ are the masses of the
components of binary and the third body, the parameters of the third
body star are calculated and are listed in the lower part of table
3. The relationship between the mass of the third body $m_{3}$ and
its orbital inclination $i_{3}$ is displayed in the left diagram of
figure 4. It is found that the minimal mass of the third companion
is $2.34(\pm0.27) \,M_{\odot}$, and this body is orbiting the binary
at a distance closer than $17.77(\pm2.55)$ AU. The eccentricity of
$e_{3} = 0.602 (\pm0.076)$ indicating the third component is
orbiting WY Per in an eccentric orbit. Using the formula given by
Mayer (1990),
\begin{equation}
 K_{RV}=\frac{2\pi}{P_{3}}\frac{a_{12}\sin i_{3}}{\sqrt{1-e_{3}^{2}}}
\end{equation}
\noindent where $K_{RV}$, $P_{3}$, $a_{12}$ are in kilometer per
second, years and AU, respectively, and considering the simplest
situation of $i_{3} = 90^{\circ}$, the semi-amplitude of the system
velocity accompanied by the light-time effect is approximately
calculated to be 1.78 $\textrm{km}~\textrm{s}^{-1}$, which is large
enough for present-day spectroscopic observations to be reliably
resolved.

The lowest mass of the third companion is $2.34\,M_{\odot}$; i.e.,
larger than the mass of the secondary component. If the tertiary
component were a normal star, we would see its spectral lines not
changing with the orbital phase of the binary, and it should be very
luminous and should contribute a large amount of third light to the
total system, unless it is not a normal main-sequence star.
Therefore, we suspect that the third body is a compact star (e.g., a
candidate black hole), it may play an important roles in the
evolution of this system. The situation resembles that in the triple
systems V Pup (Qian et al. 2008b) and WW Dra (Liao \& Qian 2010).
Certainly, such a claim must be made with considerable caution
because of two possible reasons: (1) according to Allen's tables
(Drilling \& Landolt 2000), the third companion is estimated to be a
$\sim$ A2-4 star. Actually, it is difficult to find sufficient
spectral lines to determine radial velocity of A type stars because
their rapid rotational velocity makes them too broad and weak to be
accurately measured, or (2) no detailed analysis, neither
photometric nor spectroscopic one was performed, so we did not
detect the presence of the third component. A continuous
photometric, spectroscopic as well as astrometric monitoring are
urgent to check this hypothesis in the future.
\subsection[]{Explanations of the Cyclic Period Changes in RW Leo}
For RW Leo, with equations (5) and (6), the variation of the
quadruple moment $\Delta Q$ required to reproduce the two cyclic
changes are calculated to be $\Delta Q_{1}$ = 3.52 $\times 10^{49}$
g $\textrm{cm}^{2}$ and $\Delta Q_{2}$ = 1.76 $\times 10^{49}$ g
$\textrm{cm}^{2}$ for the cyclic variation of 77.8 yr, and $\Delta
Q_{1}$ = 4.67 $\times 10^{49}$ g $\textrm{cm}^{2}$ and $\Delta
Q_{2}$ = 2.33 $\times 10^{49}$ g $\textrm{cm}^{2}$ for the cyclic
variation of 39.1 yr. Assuming conservation of the orbital angular
momentum, the total $\Delta Q$ is on the order of $10^{51}-10^{52}$
g $\textrm{cm}^{2}$ (Lanza \& Rodon\`{o} 1999), which indicates the
values of $\Delta Q_{1}$ and $\Delta Q_{2}$ for RW Leo are not
typical ones for the close binaries, suggesting, that the mechanism
of Applegate cannot interpret the two cyclical period variations of
RW Leo.

Therefore, the two cyclic period changes of RW Leo are explained as
light-travel time effect (LTTE) via the presence of the third and
fourth body, respectively. However, it is a problem that the orbits
of two additional stars do not seem to be dynamically stable,
because their period ratio is only 2. The double sinusoidal solution
may be explained in three possible ways: (1) maybe the simple
sinusoidal oscillation of $P_{4} = 39.1$ yr and $A_{4} =
0^{\textrm{d}}.022$ is caused by other mechanisms, e.g., apsidal
motion, (2) orbit-rotation resonances of two companion stars in the
system is another possible cause . This situation resembles that in
Neptune and Pluto, they rotate around the sun with the periods of
the simple ratio 3:2 ( i.e., they are 3:2 resonance), and the
similar case is $\nu$ Oct, where the periods have the simple ratio
5:2 (Ramm, D. J. et al. 2009), or (3) certainly, the double
sinusoidal solution for RW Leo is maybe merely coincidental, or, due
to the large gaps between the oldest Astronomische Nachrichten data
and Whitney's data. More observations are needed to check the
pattern of orbital period changes of RW Leo.

Whatever the causes of the two cyclical changes, the longer one can
be explained as light-travel time effect (LTTE) via the presence of
a third body. Using the equation (7) and the same absolute parameter
as Qian (2003) used ($m_{total} = 4.2M_{\odot}$), the parameters of
the third body are calculated and listed in the lower part of table
4. As shown in the right diagram of figure 4, the mass of the third
body is no less than $0.93M_{\odot}$. If this component is
main-sequence star, the computed mass would correspond to spectral
type G3-5. There is a possibility to see the spectral line of the
third body in the spectrum; also, this companion star should
contribute to the total luminosity of the system.

\section*{acknowledgements}
\small{This work is partly supported by Chinese Natural Science
Foundation (No.10973037, No. 10903026 and No.10778718), the National
Key Fundamental Research Project through grant 2007CB815406, the
Yunnan Natural Science Foundation (No. 2008CD157). We are indebted
to the many observers, amateur and professional, who obtained the
wealth of data on these two eclipsing binaries listed in table 1 and
2. Thanks are due to anonymous referee who gave us useful
suggestions and comments.}


\clearpage
\begin{table}
\tiny
\begin{center}
\caption{All available times of light minimum for WY
Per.}\label{table 1}
\begin{tabular}{lcclllll}
\hline\hline JD.Hel.  &  Min.  &  Method  &  E  &  $(O-C)_{1}$ & $(O-C)_{2}$ & Residuals & Ref. \\
2400000+   &    &     &          &     days    &     days    & days
&\\\hline
21541.34800 &I & v &   -7352 &0.0066  &0.05187 &0.00405   &(1)\\
21664.45200 &I & v &   -7315 &0.007   &0.05216 &0.00344   &(1)\\
21684.41200 &I & v &   -7309 &0.0043  &0.04945 &0.00058   &(1)\\
21817.49500 &I & v &   -7269 &0.0024  &0.04744 &-0.00239  &(1)\\
21827.48800 &I & v &   -7266 &0.014   &0.05903 &0.00914   &(1)\\
21857.43000 &I & v &   -7257 &0.0119  &0.0569  &0.0068    &(1)\\
21877.38600 &I & v &   -7251 &0.0052  &0.05019 &-0.00006  &(1)\\
21897.35000 &I & v &   -7245 &0.0064  &0.05137 &0.00099   &(1)\\
22609.35500 &I & v &   -7031 &0.0071  &0.05148 &-0.00337  &(1)\\
22619.33000 &I & v &   -7028 &0.0007  &0.04507 &-0.00984  &(1)\\
24492.50600 &I & v &   -6465 &0.0065  &0.04931 &-0.01071  &(2)\\
24795.27200 &I & v &   -6374 &0.0043  &0.04686 &-0.01242  &(2)\\
24908.40500 &I & v &   -6340 &0.0151  &0.05757 &-0.00126  &(2)\\
25234.46000 &I & v &   -6242 &0.0121  &0.0543  &-0.00262  &(3)\\
25234.46100 &I & v &   -6242 &0.0131  &0.0553  &-0.00162  &(2)\\
25244.44200 &I & v &   -6239 &0.0127  &0.05489 &-0.00196  &(2)\\
25510.61700 &I & v &   -6159 &0.0179  &0.05987 &0.00544   &(2)\\
25590.46800 &I & v &   -6135 &0.0179  &0.0603  &0.00186   &(2)\\
25620.41000 &I & v &   -6126 &0.0158  &0.05768 &0.0045    &(2)\\
25650.34600 &I & v &   -6117 &0.0077  &0.04955 &-0.00325  &(2)\\
25670.31600 &I & v &   -6111 &0.015   &0.05683 &0.00429   &(2)\\
25680.29800 &I & v &   -6108 &0.0156  &0.05743 &0.00501   &(2)\\
25700.24900 &I & v &   -6102 &0.0038  &0.04561 &-0.00654  &(2)\\
25966.42600 &I & v &   -6022 &0.011   &0.05259 &0.00456   &(2)\\
36210.55300 &I & pg&  -2943  &-0.0737 &-0.04062&0.0075    &(4)\\
36230.51500 &I & pg&  -2937  &-0.0744 &-0.04134&0.00667   &(4)\\
36446.77500 &I & pg&  -2872  &-0.0774 &-0.04452&0.00222   &(4)\\
36536.60500 &I & pg&  -2845  &-0.0798 &-0.047  &-0.00080  &(4)\\
41650.43000 &I & v &  -1308  &-0.0428 &-0.01425&-0.00848  &(5)\\
42076.31200 &I & v &   -1180 &-0.0326 &-0.0044 &-0.00252  &(6)\\
42402.36600 &I & v  &  -1082 &-0.0366 &-0.00867&-0.00982  &(7)\\
42402.37300 &I & v &   -1082 &-0.0296 &-0.00167&-0.00282  &(8)\\
42452.30000 &I & v &   -1067 &-0.0095 &0.01839 &0.01677   &(9)\\
42778.33600 &I &v  &   -969  &-0.0315 &-0.00388&-0.00856  &(10)\\
43014.56800 &I &v  &   -898  &-0.0252 &0.00222 &-0.00467  &(11)\\
43380.56300 &I &v  &   -788  &-0.0138 &0.01332 &0.00301   &(12)\\
43480.37900 &I &v  &   -758  &-0.0115 &0.01553 &0.0043    &(13)\\
43510.34600 &I &v  &   -749  &0.0114  &0.03841 &0.0269    &(13)\\
43756.53900 &I &v  &   -675  &-0.0027 &0.0241  &0.01031   &(14)\\
43916.23200 &I &v  &   -627  &-0.0116 &0.01507 &-0.00020  &(15)\\
44082.58600 &I &v  &   -577  &-0.0137 &0.01283 &-0.00398  &(16)\\
44122.50900 &I &v  &   -565  &-0.0162 &0.0103  &-0.00688  &(17)\\
44458.56300 &I &v  &   -464  &-0.0016 &0.02462 &0.00432   &(18)\\
44528.41500 &I &v  &   -443  &-0.0192 &0.00696 &-0.01399  &(19)\\
44844.50500 &I &v  &   -348  &-0.0059 &0.02    &-0.00389  &(20)\\
44984.24800 &I &v  &   -306  &-0.0021 &0.02368 &-0.0015   &(21)\\
45320.24700 &I &v  &   -205  &-0.0425 &-0.017  &-0.04528  &(22)\\
45586.46200 &I &v  &   -125  &0.0027  &0.02798 &-0.00271  &(23)\\
45649.68200 &I &v  &   -106  &0.0073  &0.03253 &0.00127   &(24)\\
46002.36000 &I &v  &   0     &0.0103  &0.03524 &0.00087   &(25)\\
47822.32700 &I &v  &   547   &0.041   &0.06443 &0.01531   &(26)\\
48564.28300 &I &v  &   770   &0.0486  &0.07141 &0.01741   &(27)\\
48677.39400 &I &v  &   804   &0.0374  &0.06012 &0.00546   &(28)\\
49043.37500 &I &v  &   914   &0.0349  &0.05731 &0.00073   &(29)\\
50141.34400 &I &v  &   1244  &0.0533  &0.0748  &0.01471   &(30)\\
50487.36300 &I &v  &   1348  &0.0515  &0.07271 &0.01257   &(31)\\
50863.32310 &I &CCD&   1461  & 0.0467 &0.0676  &0.0082    &(32)\\
51139.46300 &I &v  &   1544  & 0.0354 &0.05607 &-0.00213  &(33)\\
51279.18200 &I &CCD&   1586  & 0.0152 &0.03575 &-0.02158  &(34)\\
51535.39430 &I &CCD&   1663  & 0.0391 &0.05944 &0.00421   &(35)\\
51841.49200 &I &v  &   1755  & 0.0414 &0.06149 &0.00985   &(36)\\
51951.28230 &I &CCD&   1788  & 0.0367 &0.05669 &0.00669   &(37)\\
52267.35400 &I &v  &   1883  & 0.0317 &0.03483 &-0.00936  &(38)\\
52267.33740 &I &v  &   1883  & 0.0151 &0.05143 &0.00724   &(38)\\
52267.35400 &I &v  &   1883  & 0.0317 &0.05143 &0.00724   &(38)\\
52533.51600 &I &v  &   1963  & 0.0239 &0.04341 &0.00551   &(39)\\
52546.81890 &I &CCD&   1967  & 0.0183 &0.0378  &0.00025   &(40)\\
\hline
\end{tabular}
\tiny
\end{center}
\tiny
\end{table}

\addtocounter{table}{-1}
\begin{table}
\begin{center}
\scriptsize \caption{-continued.} \label{Table 1} \tiny \centering
\begin{tabular}{lcclllll}
\hline\hline JD.Hel.  &  Min.  &  Method  &  E  &  $(O-C)_{1}$ & $(O-C)_{2}$ & Residuals & Ref. \\
2400000+   &    &     &          &     days    &     days    & days
&\\\hline
52879.52400 &I &v  &   2067  & 0.0111 &0.03032 &0.00241   &(41)\\
52949.38870 &I &CCD&   2088  & 0.0062 &0.02537 &-0.00033  &(42)\\
52949.38970 &I &CCD&   2088  & 0.0072 &0.02637 &0.00067   &(43)\\
53002.61600 &I &CCD&   2104  & -0.0005&0.01862 &-0.00534  &(44)\\
54147.11840 &I &CCD&   2448  &-0.0284 &-0.01023&0.00383   &(45)\\
55195.13297 &I &CCD&   2763  &-0.0576 &-0.0403 &-0.0028   &(45)\\
\hline
\end{tabular}\\
\end{center}
\tiny References:(1) Hoffmeister (1921); (2) Nijland (1931); (3)
Nijland (1928); (4) Whitney (1959); (5) Peter (1972); (6) Locher
(1974); (7) Diethelm (1975); (8) Locher (1975a); (9) Locher (1975b);
(10) Peter (1976); (11) Locher (1976); (12) Locher (1977); (13)
Locher (1978a); (14) Locher (1978b); (15) Locher (1979a); (16)
Locher (1979b); (17) Locher (1979c); (18) Locher (1980a); (19)
Locher (1980a); (20) Locher (1981); (21) Locher (1982); (22)
Mavrofridis (1983); (23) Locher (1983a); (24) Locher (1983b); (25)
Locher (1984); (26) Peter (1990); (27) Locher (1992); (28) Peter
(1992); (29) Peter (1993); (30) Peter (1996); (31) Peter (1997);
(32) Diethelm (1998); (33) Guilbaut (2000); (34) O-C gateway:
http://var.astro.cz/ocgate/; (35) Safar \& Zejda (2002); (36) Locher
(2001); (37) Diethelm (2001); (38) Br\'{a}t et al. (2007); (39)
Diethelm (2003); (40) Nelson (2003); (41) Diethelm (2004); (42)
Kotkova \& Wolf (2006); (43) Zejda (2004); (44) Nelson (2004); (45)
Present paper.
\end{table}

\begin{table}
\tiny
\begin{center}
\caption{All available times of light minimum for RW
Leo.}\label{table 2}
\begin{tabular}{lccllllll}\hline\hline
JD.Hel.  &  Min.  &  Method  &  E  &  $(O-C)_{1}$ & $(O-C)_{2}$ & $(O-C)_{3}$ & Residuals & Ref. \\
2400000+   &    &     &          &   days   & days    & days  &days
&\\\hline
20960.434 &I & v &   -13292 &0.0205  &0.02083  &0.01077   &0.00577 &(1)\\
20987.352 &I & v &   -13276 &0.0179  &0.01852  &0.00798   &0.00328 &(1)\\
21251.518 &I & v &   -13119 &0.0251  &0.02808  &0.01335   &0.01115 &(1)\\
21256.556 &I & v &   -13116 &0.0155  &0.01857  &0.00371   &0.00161 &(1)\\
21342.365 &I & v &   -13065 &0.0149  &0.01873  &0.00252   &0.00122 &(1)\\
21665.400 &I & v &   -12873 &0.0022  &0.00896  &-0.01235  &-0.01055&(1)\\
21697.376 &I & v &   -12854 &0.0099  &0.01694  &-0.00486  &-0.00276&(1)\\
22412.455 &I & v &   -12429 &0.0094  &0.02266  &-0.00979  &-0.00109&(1)\\
22429.276 &I & v &   -12419 &0.0050  &0.01845  &-0.01429  &-0.00539&(1)\\
23514.516 &I & v &   -11774 &0.0066  &0.02768  &-0.01873  &-0.00163&(2)\\
24150.513 &I & v &   -11396 &0.0034  &0.02725  &-0.02216  &-0.00196&(2)\\
24261.561 &I & v &   -11330 &0.0037  &0.02789  &-0.0213   &-0.0007  &(2)\\
24584.603 &I & v &   -11138 &-0.0020 &0.02282  &-0.02423  &-0.00283&(2)\\
24993.455 &I & v &   -10895 &-0.0072 &0.0179   &-0.024    &-0.0021 &(3)\\
24993.459 &I & v &   -10895 &-0.0032 &0.0219   &-0.02     &0.0019  &(2)\\
25326.601 &I & v &   -10697 &-0.0042 &0.02053  &-0.01591  &0.00579 &(2)\\
25375.389 &I & v &   -10668 &-0.0099 &0.0148   &-0.02087  &0.00083  &(2)\\
26038.300 &I & v &   -10274 &-0.0197 &0.00285  &-0.02186  &-0.00196&(2)\\
35241.788 &I & pg&   -4804  &-0.0264 &-0.0335  &0.0013    &-0.0036 &(4)\\
35549.695 &I & pg&   -4621  &-0.0243 &-0.02867 &0.00319   &0.00119 &(4)\\
35623.725 &I & pg&   -4577  &-0.0260 &-0.0297  &0.00143   &0.00013  &(4)\\
36249.624 &I & pg&   -4205  &-0.0320 &-0.03003 &-0.00519  &-0.00049 &(4)\\
40698.254 &I & v &   -1561  &-0.0382 &-0.02528 &-0.02084  &-0.00284&(5)\\
41751.544 &I & v &   -935   &-0.0183 &-0.01344 &-0.00454  &0.00596 &(6)\\
42160.397 &I & v &   -692   &-0.0226 &-0.02155 &-0.01033  &-0.00343&(6)\\
42424.571 &I & v &   -535   &-0.0074 &-0.00889 &0.0039    &0.0084  &(6)\\
42478.413 &I & v &   -503   &-0.0067 &-0.00872 &0.0044    &0.0084  &(6)\\
42510.375 &I & v &   -484   &-0.0130 &-0.01534 &-0.00202  &0.00168 &(6)\\
42774.544 &I & v &   -327   &-0.0028 &-0.00788 &0.0072    &0.0083  &(6)\\
42828.370 &I & v &   -295   &-0.0180 &-0.0236  &-0.0082   &-0.0076 &(6)\\
42838.464 &I & v &   -289   &-0.0193 &-0.02501 &-0.00954  &-0.00904&(6)\\
42838.476 &I & v &   -289   &-0.0073 &-0.01301 &0.00246   &0.00296 &(6)\\
43139.654 &I & v &   -110   &-0.0040 &-0.01277 &0.0046    &0.0022  &(6)\\
43161.525 &I &v  &   -97    &-0.0060 &-0.01498 &0.00251   &-0.00009&(6)\\
43188.444 &I &v  &   -81    &-0.0076 &-0.01679 &0.00080   &-0.002  &(6)\\
43188.451 &I &v  &   -81    &-0.0006 &-0.00979 &0.0078    &0.005   &(6)\\
43210.315 &I &v  &   -68    &-0.0097 &-0.01921 &-0.00139  &-0.00449&(6)\\
43457.647 &I &v  &   79     &-0.0111 &-0.02304 &-0.00381  &-0.00921&(6)\\
43957.380 &I &v  &   376    &0.0075  &-0.0092  &0.01257   &0.00267 &(6)\\
44290.520 &I &v  &   574    &0.0045  &-0.01508 &0.00801   &-0.00459&(6)\\
44650.588 &I &v  &   788    &0.0089  &-0.01356 &0.01069   &-0.00461&(6)\\
45022.433 &I &v  &   1009   &0.0126  &-0.01246 &0.01264   &-0.00506&(6)\\
45791.367 &I &v  &   1466   &0.0257  &-0.00307 &0.02206   &0.00106 &(6)\\
46770.615 &I &v  &   2048   &0.0353  &0.00532  &0.02631   &0.00461 &(6)\\
47921.478 &I &v  &   2732   &0.0409  &0.01451  &0.02521   &0.00771 &(6)\\
49055.500 &I &CCD&   3406   &0.0308  &0.01221  &0.00823   &-0.00087&(7)\\
49743.669 &I &v  &   3815   &0.0409  &0.02825  &0.01359   &0.01079 &(6)\\
50571.4637&I &CCD&   4307   &0.0258  &0.02071  &-0.00687  &-0.00167 &(6)\\
51274.763 &I &CCD&   4725   &0.0233  &0.02404  &-0.01359  &-0.00219 &(7)\\
51303.377 &I &v  &   4742   &0.0341  &0.03503  &-0.00296  &0.00864  &(6)\\
51626.4107&I &CCD&   4934   &0.0201  &0.02346  &-0.01881  &-0.00461 &(8)\\
51636.5150&I &v  &   4940   &0.0292  &0.03255  &-0.00977  &0.00443  &(9)\\
51685.3088&I &CCD&   4969   &0.0293  &0.03302  &-0.00992  &0.00468  &(8)\\
52395.3470&I &v  &   5391   &0.0355  &0.04315  &-0.00551  &0.01339  &(10)\\
52750.3650&I &v  &   5602   &0.0376  &0.04656  &-0.00185  &0.01855  &(11)\\
53029.6429&I &CCD&   5768   &0.0138  &0.02342  &-0.02299  &-0.00179 &(12)\\
53135.6407&I &CCD&   5831   &0.0116  &0.02136  &-0.0239   &-0.0025  &(13)\\
53374.5670&I &v  &   5973   &0.0172  &0.02723  &-0.01498  &0.00682  &(14)\\
53771.63309&I&CCD&   6209   &0.0038  &0.01362  &-0.02233  &-0.00053 &(15)\\
53771.63378&I&CCD&   6209   &0.0045  &0.01432  &-0.02163  &0.00017  &(16)\\
53771.63448&I&CCD&   6209   &0.0052  &0.01502  &-0.02093  &0.00087  &(15)\\
54138.4234 &I&CCD&   6427   &0.0003  &0.00943  &-0.02059  &0.00071  &(16)\\
54150.2001 &I&CCD&   6434   &-0.0008 &0.00832  &-0.02153  &-0.00023 &(17)\\
54202.3591 &I&CCD&   6465   &-0.0005 &0.0084   &-0.02056  &0.00054  &(18)\\
54207.4032 &I&pe &   6468   &-0.0040 &0.00489  &-0.02399  &-0.00289 &(19)\\
\hline
\end{tabular}
\tiny
\end{center}
\tiny
\end{table}

\addtocounter{table}{-1}
\begin{table}
\begin{center}
\scriptsize \caption{-continued.} \label{Table 2} \tiny \centering
\begin{tabular}{lccllllll}\hline\hline
JD.Hel.  &  Min.  &  Method  &  E  &  $(O-C)_{1}$ & $(O-C)_{2}$ & $(O-C)_{3}$ & Residuals & Ref. \\
2400000+   &    &     &          &   days   & days    & days  &days
&\\\hline
54207.4065 &I&pe &   6468   &-0.0007 &0.00819  &-0.02069  &0.00041  &(19)\\
54207.40717&I&CCD&   6468   &0.0000  &0.00889  &-0.01999  &0.00111  &(20)\\
54488.38902&I&CCD&   6635   &-0.0024 &0.00545  &-0.01908  &0.00112  &(21)\\
54939.3062 &I&CCD&   6903   &-0.0060 &-0.00039 &-0.01835  &-0.00015 &(16)\\
55215.24169&I&CCD&   7067   &-0.0071 &-0.00324 &-0.0172   &-0.0006  &(21)\\
\hline
\end{tabular}\\
\end{center}
 \tiny References:(1) Hoffmeister (1921); (2) Nijland (1931); (3) Nijland (1928); (4) Whitney (1959); (5) Silhan (1971);
 (6) refer to timings from the EBMD by internet;
 (7) O-C gateway: http://var.astro.cz/ocgate/; (8) Zejda
(2002); (9) BBSAG observers (2000); (10) BBSAG observers (2002);
(11) Diethelm (2003); (12) Zejda (2004); (13) Dvorak (2005); (14)
Locher (200105); (15) Br\'{a}t et al. (2007); (16) Dogru et al.
(2009); (17) VSOLJ(No. 46); (18) Borkovits et al. (2008); (19)
Hubscher (2007); (20)Br\'{a}t et al. (2009); (21) Present paper.
\end{table}

\begin{table}
\caption{Parameters of the orbit of the potential third body for WY
Per.}\label{table 3}
\begin{center}
\small
\begin{tabular}{ll}\hline\hline
 Parameters    & Values  \\\hline
$JD_{0}$(HJD) &2446002.3248$\pm0.0053$ \\
$P_{0}$(day) &3.3271230$\pm0.00000397$\\
$P_{3}$(yr) & 71.5$\pm1.9$\\
$\omega_{3}(^{{\circ}})$ & 164.6$\pm12.1$\\
$\beta$(day $\textrm{cycle}^{-1}$)&0(assumed)\\
$e_{3}$  &0.602$\pm0.076$\\
$A$(day)  & 0.0739$\pm0.0062$\\
$T_{3}$(HJD)& 2453132.3$\pm290.9$\\
\hline
$a_{12}\sin{i_{3}}$(AU)& 12.80$\pm1.07$\\
$f(m)(M_{\odot})$  & 0.41$\pm0.10$\\
$m_{3min}(M_{\odot})$&2.34$\pm0.27$\\
$a_{3max}$(AU)  &17.77$\pm2.55$\\
\hline
\end{tabular}
\end{center}
\end{table}

\begin{table}
\caption{Parameters of the orbit of the potential third body for RW
Leo.}\label{table 4}
\begin{center}
\small
\begin{tabular}{ll}\hline\hline
 Parameters    & Values  \\\hline
$JD_{0}$(HJD) &2443324.7439$\pm0.0047$\\
$P_{0}$(day) &1.68254106$\pm0.00000068$\\
$P_{3}$(yr) & 77.8$\pm1.5$\\
$\omega_{3}(^{{\circ}})$ & 126.6$\pm14.6$\\
$\beta$(day $\textrm{cycle}^{-1}$)&0(assumed)\\
$e_{3}$  &0.732$\pm0.065$\\
$A$(day)  & 0.033$\pm0.004$\\
$T_{3}$(HJD)&2452896.4$\pm250.0$\\
\hline
$a_{12}\sin{i_{3}}$(AU)& 5.72$\pm0.69$\\
$f(m)(M_{\odot})$  & 0.031$\pm0.011$\\
$m_{3min}(M_{\odot})$&0.93$\pm0.13$\\
$a_{3max}$(AU)  &25.72$\pm4.72$\\
\hline
\end{tabular}
\end{center}
\end{table}
\clearpage
\begin{figure}
\centering
\includegraphics[width=9.0cm]{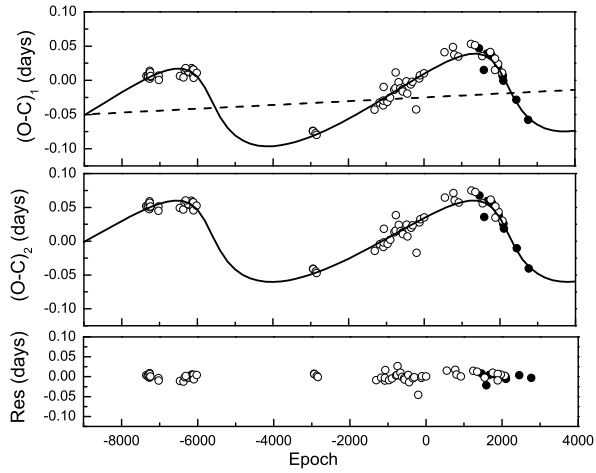}
\caption{$(O-C)$ diagram of WY Per calculated with equations (2) and
(3) based on all available times of light minimum. The upper panel :
$(O-C)_1$ diagram of WY Per computed with equation (1). The open
circles refer to photographic or visual observations, filled circles
refer to CCD ones. The solid line refers to a combination of a
linear ephemeris and a cyclical period variation with an
eccentricity of 0.602, and the dashed line to a new linear
ephemeris. The middle panel : $(O-C)_2$ curve of WY Per as described
by the last term in equation (2) (solid line), after removing the
linear correction term. The symbols are the same as in the upper
panel. The lower panel : The residuals from fit with equations (2)
and (3). The symbols are the same as in the upper panel.}
\label{figure 1}
\end{figure}
\begin{figure}
\centering
\includegraphics[width=9.0cm]{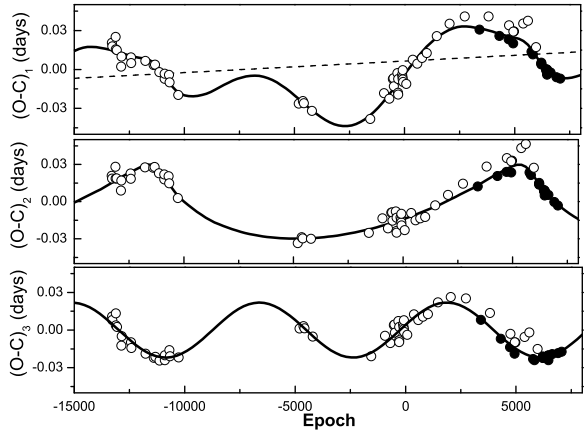}
\caption{$(O-C)$ diagram of RW Leo. The upper panel : $(O-C)_1$
diagram of RW Leo computed with equation (4). The open circles refer
to photographic or visual observations, filled circles refer to CCD
or photoelectric ones. The solid line refers to a combination of a
linear, a cyclical period variation with an eccentricity of 0.732
and a pure cyclical period variation, the dashed line to a new
linear ephemeris. The middle panel : $(O-C)_2$ curve of RW Leo as
described by the fourth term in equation (2) (solid line), after
removing the other terms. The symbols are the same as in the upper
panel. The lower panel : $(O-C)_3$ curve of RW Leo as described by
the sinusoidal term $A_{4}\sin(\omega_{4} E-\phi)$ (solid line). The
symbols are the same as in the upper panel.} \label{figure 2}
\end{figure}
\clearpage
\begin{figure}
\centering
\includegraphics[width=9.0cm]{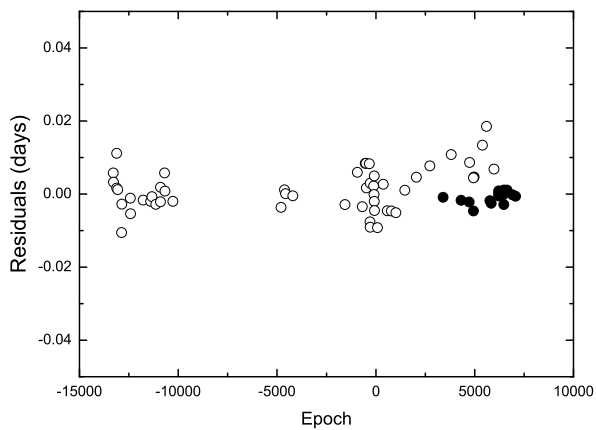}
\caption{The residuals from fit for RW Leo. The symbols are the same
as in the figure 2.} \label{figure 3}
\end{figure}

\begin{figure}
\centering
\includegraphics[width=9.0cm]{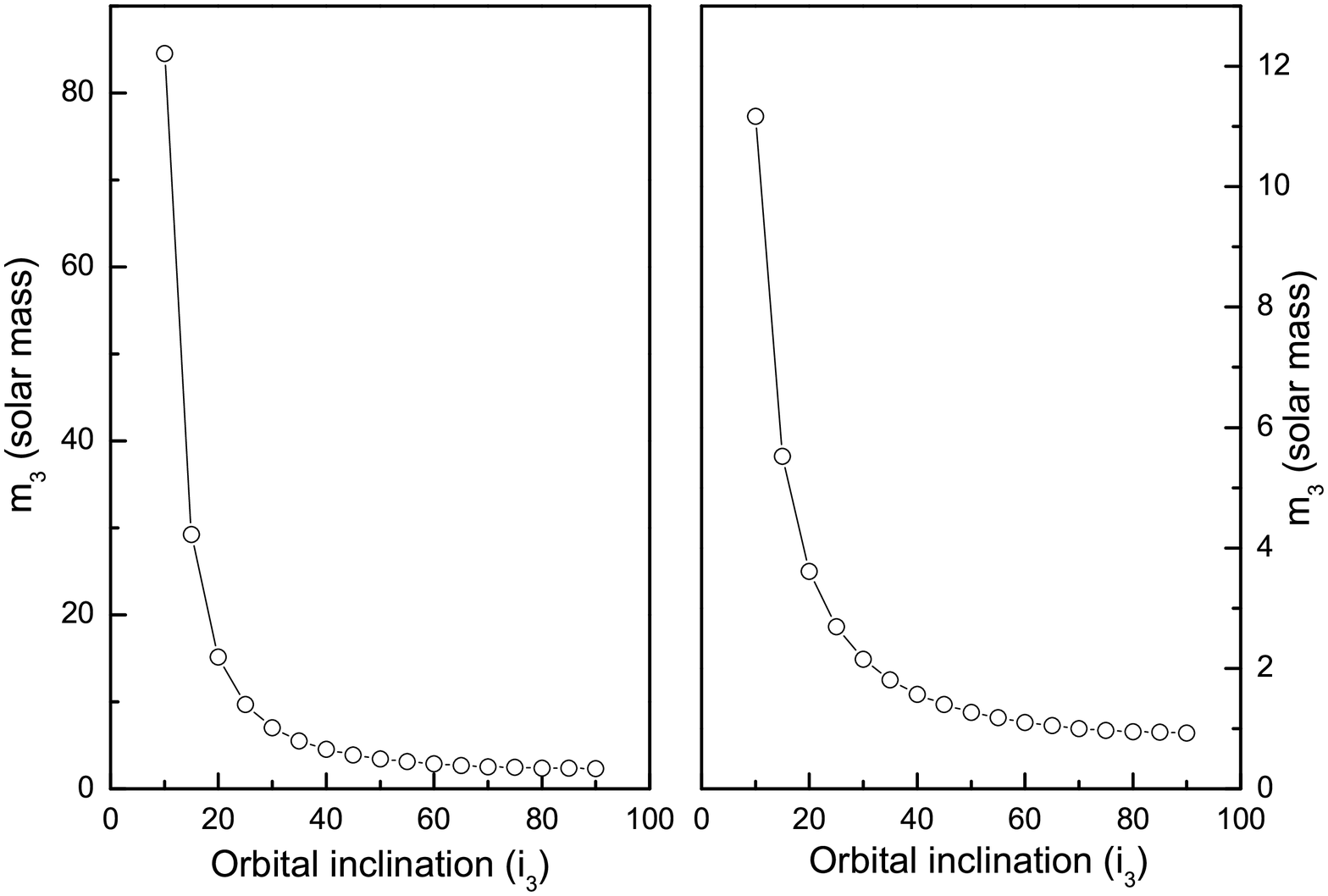}
\caption{Relations between the mass and the orbital inclinations of
the third component. The left diagram is for that of WY Per, and the
right one for that of RW Leo.} \label{figure 4}
\end{figure}

\begin{thebibliography}{}
\bibitem{}
Applegate J. H. 1992, ApJ, 385, 621
\bibitem{}
BBSAG observers. 2000, Bull. BBSAG, 122
\bibitem{}
BBSAG observers. 2002, Bull. BBSAG, 128
\bibitem{}
Borkovits, T., \& Heged\"{u}s, T. 1996, A\&AS, 120,63
\bibitem{}
Borkovits, T., van Cauteren, P., Lampens, P., Dufoer, S., Kleidis,
S., van Leenhove, M., Csizmadia, S., Regaly, Z., Patkos, L.,
Klagyivik, P., and 3 coauthors. 2008, Inf. Bull. Var. Stars, 5835
\bibitem{}
Brancewicz, H. K., \& Dworak, T. Z. 1980, AcA, 30, 501
\bibitem{}
Br\'{a}t, L., Zejda, M., \& Svoboda, P. 2007, OEJV, 74, 1
\bibitem{}
Br\'{a}t, L., Trnka, J., Lehky, M., Smelcer, L., Kucakova, H.,
Ehrenberger, R., Dreveny, R., Lomoz, F., Marek, P., Kocian, R., and
13 coauthors. 2009, OEJV, 107, 1
\bibitem{}
Budding, E. 1984, BICDS, 27, 91
\bibitem{}
Budding, E., Erdem, A., \c{C}i\c{c}ek, C., Bulut, I., Soydugan, F.,
Soydugan, E., Baki\c{s}, V., \& Demircan, O. 2004, A\&A, 417, 263
\bibitem{}
Chambliss, C. R. 1992, PASP, 104, 663
\bibitem{ }
Diethelm, R. 1975, Bull. BBSAG, 19
\bibitem{}
Diethelm, R. 1998, Bull. BBSAG, 117
\bibitem{}
Diethelm, R. 2001, Bull. BBSAG, 124
\bibitem{}
Diethelm, R. 2003, Inf. Bull. Var. Stars, 5438
\bibitem{}
Diethelm, R. 2004, Inf. Bull. Var. Stars, 5543
\bibitem{}
Dogru, S. S., Erdem, A., Donmez, A., Bulut, A., Akin, T., Dogru, D.,
Cicek, C., \& Soydugan, F. 2009, Inf. Bull. Var. Stars, 5893
\bibitem{}
Drilling, J. S., Landolt, A. U., 2000, in Allen's Astrophysical
Quantities, ed. A. N. Cox (4th ed.; New York: Springer), 388
\bibitem{}
Dvorak, S. W. 2005, Inf. Bull. Var. Stars, 5603
\bibitem{}
Frieboes-Conde, H., \& Herczeg, T. 1973, A\&AS, 12, 1
\bibitem{}
Guilbaut, P. 2000, Bull. BBSAG, 123
\bibitem{}
Hall, D. S. 1989, Space Sci. Rev, 50, 219
\bibitem{}
Halbedel, E. M. 1984, PASP, 96, 98
\bibitem{}
Hoffmeister, Cu  1921, AN, 214, 1
\bibitem{}
Hubscher, J. 2007, Inf. Bull. Var. Stars, 5802
\bibitem{}
Kaitchuck, R. H., Honeycutt, R. K., \& Schlegel, E. M. 1985, PASP,
97, 1178
\bibitem{}
Koch, R. H., Wood, F. B., Florkowski, D. R., \& Oliver, J. P. 1979,
Inf. Bull. Var. Stars, 1709
\bibitem{}
Kotkova, L., \& Wolf, M. 2006, Inf. Bull. Var. Stars, 5676
\bibitem{}
Kreiner J. M., Kim C.H., \& Nha I.S. 2001, An Altas of O-C Diagrams
of Eclipsing Binary Star. Wydawnictwo Naukowe Akademii
Pedagogicznej, Cracow, Poland
\bibitem{}
Lanza, A. F. 2006, MNRAS, 369, 1773
\bibitem{}
Lanza, A.F., \& Rodon\`{o}, M. 1999, A\&A, 349, 887
\bibitem{}
Lanza, A.F., \& Rodon\`{o}, M. 2002, AN, 323, 424
\bibitem{}
Liao W.-P., \& Qian S.-B. 2010, MNRAS, in press,
doi:10.1111/j.1365-2966.2010.16584.x
\bibitem{}
Locher, K. 1974, Bull. BBSAG, 13
\bibitem{}
Locher, K. 1975a, Bull. BBSAG, 19
\bibitem{}
Locher, K. 1975b, Bull. BBSAG, 21
\bibitem{}
Locher, K. 1976, Bull. BBSAG, 29
\bibitem{}
Locher, K. 1977, Bull. BBSAG, 34
\bibitem{}
Locher, K. 1978a, Bull. BBSAG, 36
\bibitem{}
Locher, K. 1978b, Bull. BBSAG, 39
\bibitem{}
Locher, K. 1979a, Bull. BBSAG, 42
\bibitem{}
Locher, K. 1979b, Bull. BBSAG, 44
\bibitem{}
Locher, K. 1979c, Bull. BBSAG, 45
\bibitem{}
Locher, K. 1980a, Bull. BBSAG, 49
\bibitem{}
Locher, K. 1980b, Bull. BBSAG, 51
\bibitem{}
Locher, K. 1981, Bull. BBSAG, 56
\bibitem{}
Locher, K. 1982, Bull. BBSAG, 58
\bibitem{}
Locher, K. 1983a, Bull. BBSAG, 68
\bibitem{}
Locher, K. 1983b, Bull. BBSAG, 69
\bibitem{}
Locher, K. 1984, Bull. BBSAG, 74
\bibitem{}
Locher, K. 1992, Bull. BBSAG, 99
\bibitem{}
Locher, K. 2001, Bull. BBSAG, 124
\bibitem{}
Locher, K. 2005, OEJV, 3
\bibitem{}
Mavrofridis, G. 1983, Bull. BBSAG, 64
\bibitem{} Mayer, P., 1990, BAICz, 41, 231
\bibitem{}
Nelson, R. H. 2003, Inf. Bull. Var. Stars, 5371
\bibitem{}
Nelson, R. H. 2004, Inf. Bull. Var. Stars, 5493
\bibitem{}
Nijland, A. A. 1931, AN, 242, 9
\bibitem{}
Nijland, A. A. 1928, Acta Astr. Ser. c, 1, 29
\bibitem{}
Peter, H. 1972, Bull. BBSAG, 6
\bibitem{}
Peter, H. 1976, Bull. BBSAG, 25
\bibitem{}
Peter, H. 1990, Bull. BBSAG, 93
\bibitem{}
Peter, H. 1992, Bull. BBSAG, 100
\bibitem{}
Peter, H. 1993, Bull. BBSAG, 103
\bibitem{}
Peter, H. 1996, Bull. BBSAG, 111
\bibitem{}
Peter, H. 1997, Bull. BBSAG, 114
\bibitem{}
Rovithis-Livaniou, H., Kranidiotis, A. N., Rovithis, P., \&
Athanassiades, G. 2000, A\&A, 354, 904
\bibitem{}
Qian Shengbang. 2003, PASJ, 55, 289
\bibitem{}
Qian, S.-B., Dai, Z.-B., Zhu, L.-Y., Liu, L., He, J.-J., Liao,
W.-P., \& Li, L.-J. 2008a, ApJ, 689, 49
\bibitem{}
Qian, S.-B., Liao, W.-P., Fern\'{a}ndez Laj\'{u}s, E., 2008b, ApJ,
687, 466
\bibitem{}
Ramm, D. J., Pourbaix, D., Hearnshaw, J. B., \& Komonjinda, S. 2009,
MNRAS, 394, 1695
\bibitem{}
Safar, J., \& Zejda, M. 2002, Inf. Bull. Var. Stars, 5263
\bibitem{}
Silhan, J. 1971, Brno Contr. 12
\bibitem{ }
Svechnikov, M. A., \& Kuznetsova, E. F. 1990, Kat- alog
Priblizhennykh Fotometricheskikh i Absoliut- nykh Elementov
Zatmennykh Peremennykh Zvezd (Sverdlovsk: Izd-vo Ural'skogo
Universiteta)
\bibitem{}
Walker, R. L. 1992, BAAS, 24, 1138
\bibitem{}
Whitney, B. S. 1959, AJ, 64, 258
\bibitem{}
Wolf, M. 1917, AN, 205, 23
\bibitem{}
Wolf, M., Mayer, P., Zasche, P., Sarounova, L., \& Zejda, M. 2004,
ASPC, 318, 255
\bibitem{}
Zejda, M. 2002, Inf. Bull. Var. Stars, 5287
\bibitem{}
Zejda, M. 2004, Inf. Bull. Var. Stars, 5583
\end{thebibliography}
\end{document}